\documentclass[runningheads]{llncs}
\usepackage[T1]{fontenc}
\usepackage{graphicx}
\usepackage{amsmath}
\usepackage{hyperref}
\usepackage{cite}
\usepackage{booktabs}
\usepackage{float}
\usepackage{multirow}
\usepackage[table]{xcolor}
\usepackage{placeins}

\title{Synthetic-to-Real Transfer in Cerebral Microbleed Generation and Segmentation}

\makeatletter
\newcommand{\printfnsymbol}[1]{\textsuperscript{\@fnsymbol{#1}}}
\makeatother

\author{
To-Liang Hsu\orcidID{0009-0008-7173-0939} \and \\
Ting-Yu Lai\orcidID{0009-0007-3561-0211} \and \\ 
Ching-Ting Lin\orcidID{0000-0002-6925-9999} \and \\Chun-Hao Huang\orcidID{0009-0002-6663-1967}\and \\
Wei-Chun Wang\thanks{Corresponding author: \email{017141@tool.caaumed.org.tw}} \orcidID{0000-0002-2239-8174}
}

\institute{Artificial Intelligence and Robotics Innovation Center, China Medical University Hospital, China Medical University, Taichung, Taiwan \\
\email{017141@tool.caaumed.org.tw}}

\begin{document}

\titlerunning{Synthetic-to-Real Transfer in Microbleed}
\authorrunning{Hsu et al.}
\maketitle

\begin{abstract}
The development of automated cerebral microbleed (CMB) detection models is hindered by the low prevalence of CMBs and the high cost of expert annotation. To address this limitation, we developed a synthetic CMB generation pipeline and investigated the effectiveness of synthetic lesions for training deep learning detectors. Models trained solely on synthetic data achieved substantial detection performance, reaching approximately 87\% of the lesion sensitivity of their real-trained counterparts. We further investigated the complementary roles of synthetic and real data under different training paradigms, revealing that a significant performance gap remains. Moreover, we found that successful synthetic-to-real transfer is strongly dependent on the downstream detection architecture, providing new insight into both the potential and limitations of synthetic data for CMB detection.
\end{abstract}

\section{Introduction}

Cerebral microbleeds (CMBs) are small, round or ovoid chronic deposits of hemosiderin-laden macrophages associated with cerebral small vessel disease, including cerebral amyloid angiopathy and hypertensive arteriopathy. They appear as hypointense lesions on T2*-weighted gradient-recalled echo (GRE) or susceptibility-weighted imaging (SWI), and are increasingly recognized as important imaging biomarkers associated with an increased risk of intracerebral hemorrhage and ischemic stroke, cognitive impairment and dementia \cite{charidimou2011}. Clinical assessment of CMB burden is commonly performed using standardized visual rating scales such as the Microbleed Anatomical Rating Scale (MARS) and the Brain Observer MicroBleed Scale (BOMBS) \cite{gregoire2009mars,cordonnier2009bombs}, which remain labor-intensive. Automated detection of CMBs remains challenging because the lesions are small, sparsely distributed, and frequently resemble normal vascular structures or imaging artifacts, including vessel cross-sections, calcifications, and susceptibility-related signal voids. Furthermore, the relatively low prevalence of CMBs in the general population \cite{haller2018} limits the availability of positive training samples, making expert annotation both expensive and time-consuming.

To alleviate the data scarcity problem, synthetic CMB generation has emerged as a promising direction. Momeni et al. proposed an analytical generation framework that models CMBs using randomized Gaussian shapes and inserts synthetic lesions into healthy brain locations \cite{momeni2021}. They demonstrated the feasibility of training a patch-level classifier using synthetic CMBs alone. When extended to whole-brain detection, however, the synthetic-trained model produced approximately nine false positives per scan.
CenSynCMB introduced a physics-guided synthesis framework that generates both synthetic CMBs and vessel-like and calcification-like mimics in real training backgrounds \cite{he2026censyncmb}. CenSynCMB explicitly models common CMB mimics to improve false-positive control, but primarily evaluates synthetic data as an auxiliary augmentation strategy alongside real training data rather than as an independent training source.

The effectiveness of synthetic training data, however, may also depend on the downstream detection framework. Automated CMB detection methods have evolved from handcrafted image-processing techniques and conventional machine-learning classifiers toward deep learning-based detection and segmentation frameworks \cite{almasni2020,sundaresan2023,tsuchida2024}. The VALDO challenge demonstrated substantial progress in automated CMB detection while also revealing considerable variability in performance across methods and individual cases \cite{sudre2024}. Despite the diversity of modern detection approaches, previous synthetic CMB studies both evaluated synthetic data within a single downstream detection framework. Consequently, whether synthetic-to-real transfer depends on the downstream detector architecture remains largely unexplored.

\begin{figure}[htbp]
\centering
\includegraphics[width=\textwidth]{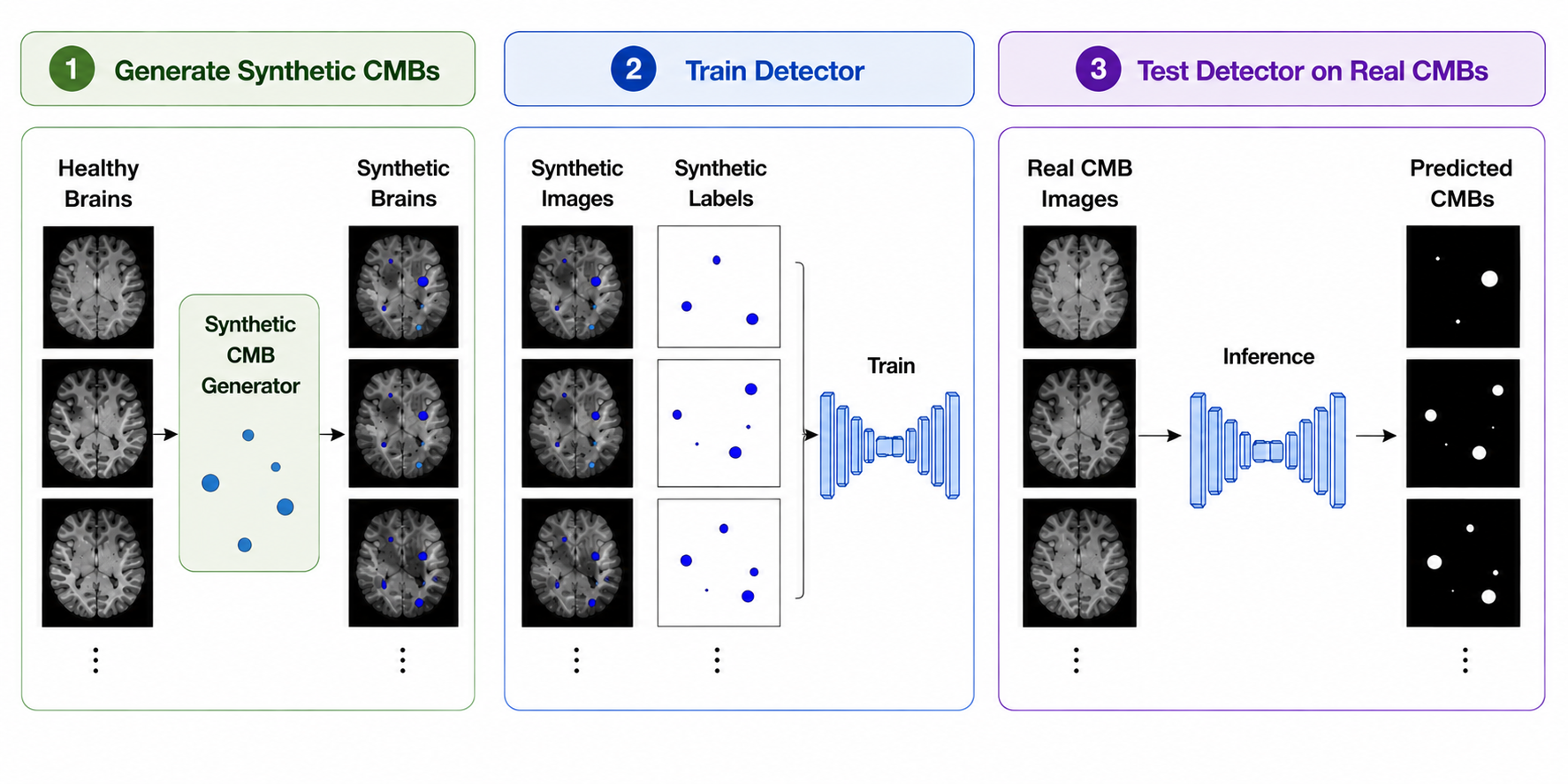} 
\caption{Overview}
\label{fig:overview}
\end{figure}

In this work, we investigate the effectiveness of synthetic CMBs as a training resource for automated CMB detection. We first developed a physics-inspired synthetic lesion generation pipeline that generates synthetic targets on CMB-negative scans, ensuring a strict separation between synthetic training data and real positive lesions. Using this framework, we systematically compare four training paradigms: pure synthetic CMB training, pure real CMB training, synthetic training with real-data validation, and mixed real-synthetic training. Furthermore, we evaluate synthetic-to-real transfer across both MONAI 3D U-Net and nnU-Net architectures to examine the influence of detector design on transferability. An overview of the synthetic-to-real detection framework is shown in Fig.~\ref{fig:overview}. Implementation details can be found on GitHub, with additional illustrations
available on the project website.\footnote{%
Project repository:
\href{https://github.com/Lukehsu1999/synthetic-to-real-cmb-detection}{GitHub};
project website:
\href{https://lukehsu1999.github.io/synthetic-to-real-cmb-detection/}{Website}.%
}

\section{Data}
\subsection{Data Acquisition}
All magnetic resonance imaging (MRI) examinations used in this study were retrospectively collected from China Medical University Hospital under Institutional Review Board approval (IRB: CMUH110-REC2-245). All examinations were acquired using susceptibility-weighted angiography (SWAN) sequences on GE Healthcare MRI systems in routine clinical practice. To reflect real-world imaging variability, data were collected from multiple scanner models, including the Discovery MR750w, Optima MR450w, Signa HDxt, Signa Architect, and Signa Voyager platforms. Both 1.5T and 3T MRI systems were included. 

The in-plane resolution of the images ranged from
$0.43 \times 0.43~\mathrm{mm}^2$
to
$0.51 \times 0.51~\mathrm{mm}^2$,
with a median resolution of
$0.47 \times 0.47~\mathrm{mm}^2$.
Slice spacing ranged from
$0.8~\mathrm{mm}$
to
$1.4~\mathrm{mm}$,
with a median of
$1.0~\mathrm{mm}$.

A total of 205 MRI examinations were included in this study, comprising 104 cerebral microbleed (CMB)-positive cases and 101 CMB-negative cases. The detailed allocation of these cases into the training, validation, and testing cohorts is described in \ref{subsec:dataset_partitioning}.

The inclusion of multiple scanner models, magnetic field strengths, and acquisition protocols reflects routine clinical practice, providing a realistic and heterogeneous dataset.

\subsection{Microbleed Annotation}
Only CMBs with diameters between $2.0$ and $10.0~\mathrm{mm}$ were included in the annotations, consistent with established CMB imaging conventions \cite{wardlaw2013strive}. All lesions were manually labeled on SWAN images by an experienced neurologist using voxel-wise segmentation masks.

Only cases with confirmed CMB lesions were assigned voxel-wise segmentation labels, while CMB-negative examinations were verified to contain no visible CMBs and were used as negative controls during model development and evaluation.

\subsection{Dataset Partitioning}
\label{subsec:dataset_partitioning}

The dataset was divided into training and hold-out test sets, with the latter reserved exclusively for final evaluation and not used during model training or model selection (Table~\ref{tab:dataset_partition}).

To maintain a comparable distribution of disease burden across the two sets, CMB-positive subjects were stratified by the number of annotated CMBs into three categories: fewer than 5 lesions, 5--9 lesions, and 10 or more lesions. Subjects were then assigned while preserving the distribution of these lesion-burden categories (Table~\ref{tab:lesion_burden}).

The 78 CMB-positive training examinations were used to train the real-data benchmark models, whereas the 75 CMB-negative training examinations served exclusively as host images for synthetic CMB generation and were not included in real-data training.

\begin{table}[htbp]
\centering

\begin{minipage}[t]{0.52\textwidth}
\centering
\caption{Dataset partition used in this study.}
\label{tab:dataset_partition}
\small
\setlength{\tabcolsep}{3pt}
\begin{tabular}{lccc}
\toprule
\textbf{Dataset} &
\textbf{CMB+} &
\textbf{CMB--} &
\textbf{Lesions} \\
\midrule
Training & 78 & 75$^{*}$ & 687 \\
Hold-out test & 26 & 26 & 171 \\
\bottomrule
\end{tabular}

\vspace{2pt}
{\scriptsize $^{*}$Synthetic hosts.}
\end{minipage}
\hfill
\begin{minipage}[t]{0.44\textwidth}
\centering
\caption{Distribution of CMB-positive cases by lesion burden.}
\label{tab:lesion_burden}
\small
\setlength{\tabcolsep}{4pt}
\begin{tabular}{lrr}
\toprule
\textbf{Burden} &
\textbf{Train} &
\textbf{Test} \\
\midrule
$<5$ CMBs      & 42 & 14 \\
5--9 CMBs      & 18 & 6 \\
$\geq10$ CMBs  & 18 & 6 \\
\midrule
Total          & 78 & 26 \\
\bottomrule
\end{tabular}
\end{minipage}

\end{table}

\section{Methods}
\subsection{Overview}
This section describes the complete methodology of the proposed framework. We first introduce the image preprocessing procedure, followed by the synthetic cerebral microbleed generation pipeline. Next, we describe the segmentation networks and training strategies used to evaluate the effectiveness of synthetic data. Finally, we present the evaluation metrics adopted for performance assessment.

\subsection{Preprocessing}

All MRI examinations underwent a standardized preprocessing pipeline prior to synthetic data generation and model training, including the real CMB-positive training set, the CMB-negative synthetic host dataset, and the independent hold-out test set.

First, the original DICOM images were converted to NIfTI format while preserving the original spatial metadata. As no pretrained brain-extraction model was available specifically for SWAN images, brain extraction was performed using a dedicated nnU-Net v2 model trained for this purpose. The model was trained on a subset of 20 delineated SWAN volumes from CMB-positive training cohort. The resulting binary brain masks were applied to the SWAN images to remove non-brain voxels.

\begin{figure}[htbp]
\centering
\includegraphics[width=\textwidth]{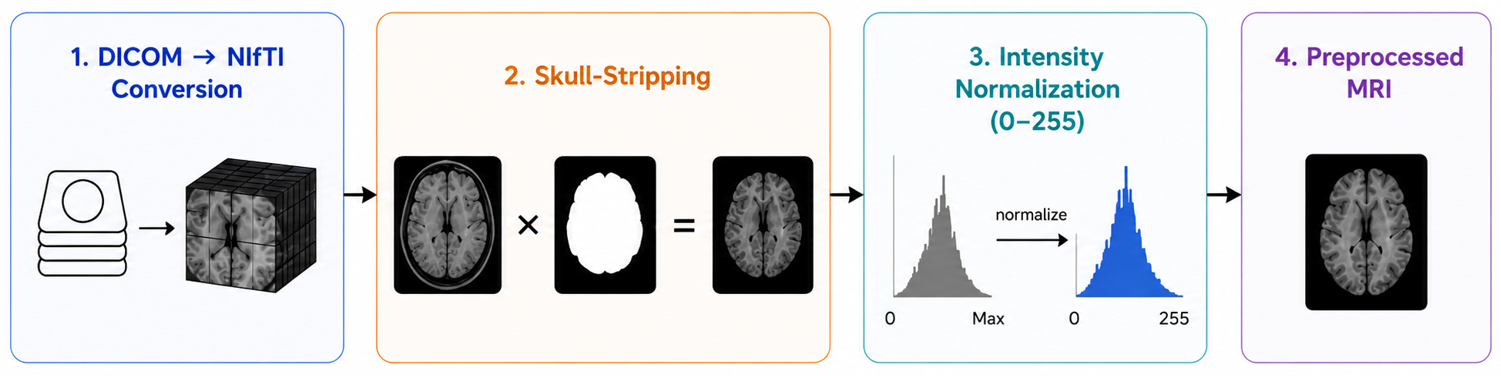} 
\caption{Preprocessing Procedure}
\label{fig:preprocessing_flow}
\end{figure}

The skull-stripped images were subsequently normalized to an 8-bit intensity range of 0--255. The common preprocessing pipeline is illustrated in Fig.~\ref{fig:preprocessing_flow}. Downstream detection frameworks may add on their own customized preprocessing after this procedure, as described in Section~\ref{subsec:detection_network}.

This standardized preprocessing procedure provided a consistent image representation across all datasets prior to synthetic lesion generation and model training.

\subsection{Synthetic Microbleed Generation Pipeline}
\subsubsection{Physical Motivation}
\begin{figure}[htbp]
\centering
\includegraphics[width=\textwidth]{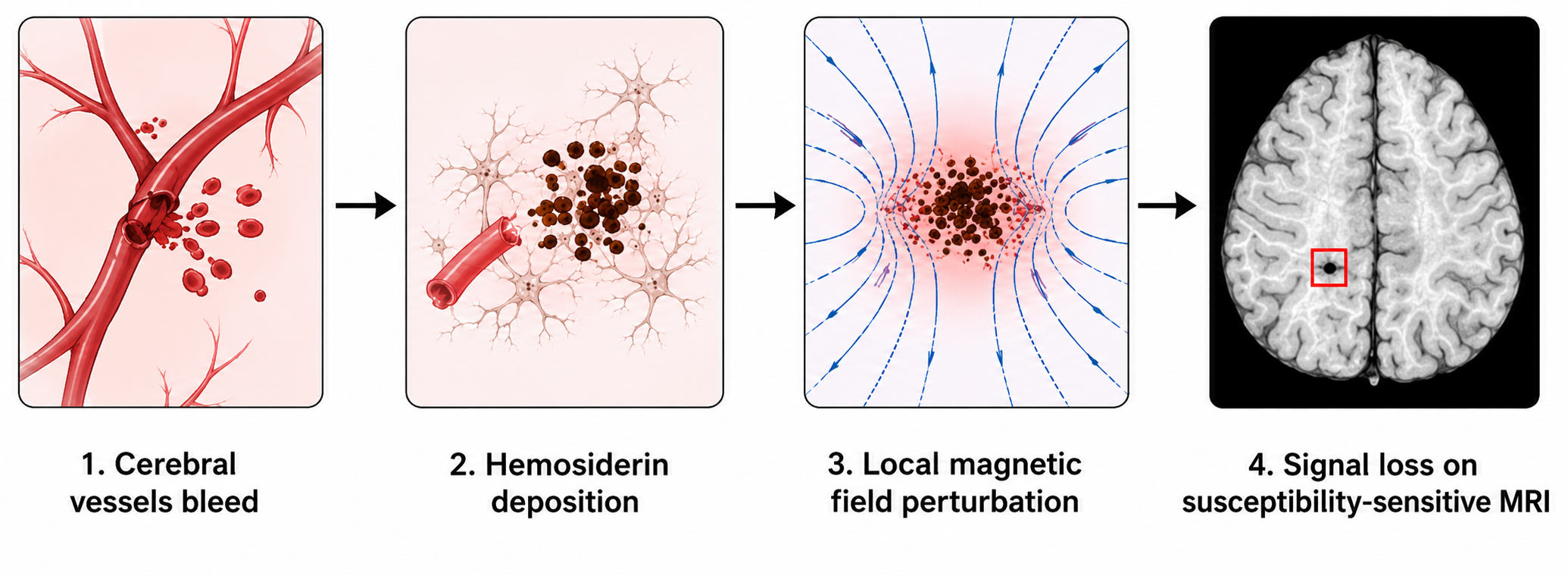} 
\caption{Physical mechanism underlying CMB appearance on susceptibility-sensitive MRI}
\label{fig:CMB_physics}
\end{figure}

Cerebral microbleeds are small chronic hemorrhagic lesions resulting from hemosiderin deposition following blood leakage from small cerebral vessels. Hemosiderin is highly paramagnetic and locally perturbs the magnetic field, leading to accelerated transverse dephasing in susceptibility-sensitive MRI sequences such as SWAN. Consequently, CMBs appear as small hypointense signal voids on SWAN images. The process is illustrated in Fig. ~\ref{fig:CMB_physics}.

Rather than explicitly simulating the underlying magnetic susceptibility effects and MRI signal formation, we focus on modeling the signal loss visual effect of CMBs on SWAN.

\subsubsection{Synthetic Lesion Generation}
Our synthetic lesion generator aims to approximate real CMB appearance while maintaining controllable generation parameters. To avoid information leakage, all real-lesion measurements used to inform the generator were obtained exclusively from the real-positive training cohort. The independent hold-out test set was never used for generator parameterization and was reserved solely for final performance evaluation.

\begin{figure}[htbp]
\centering
\includegraphics[width=\textwidth]{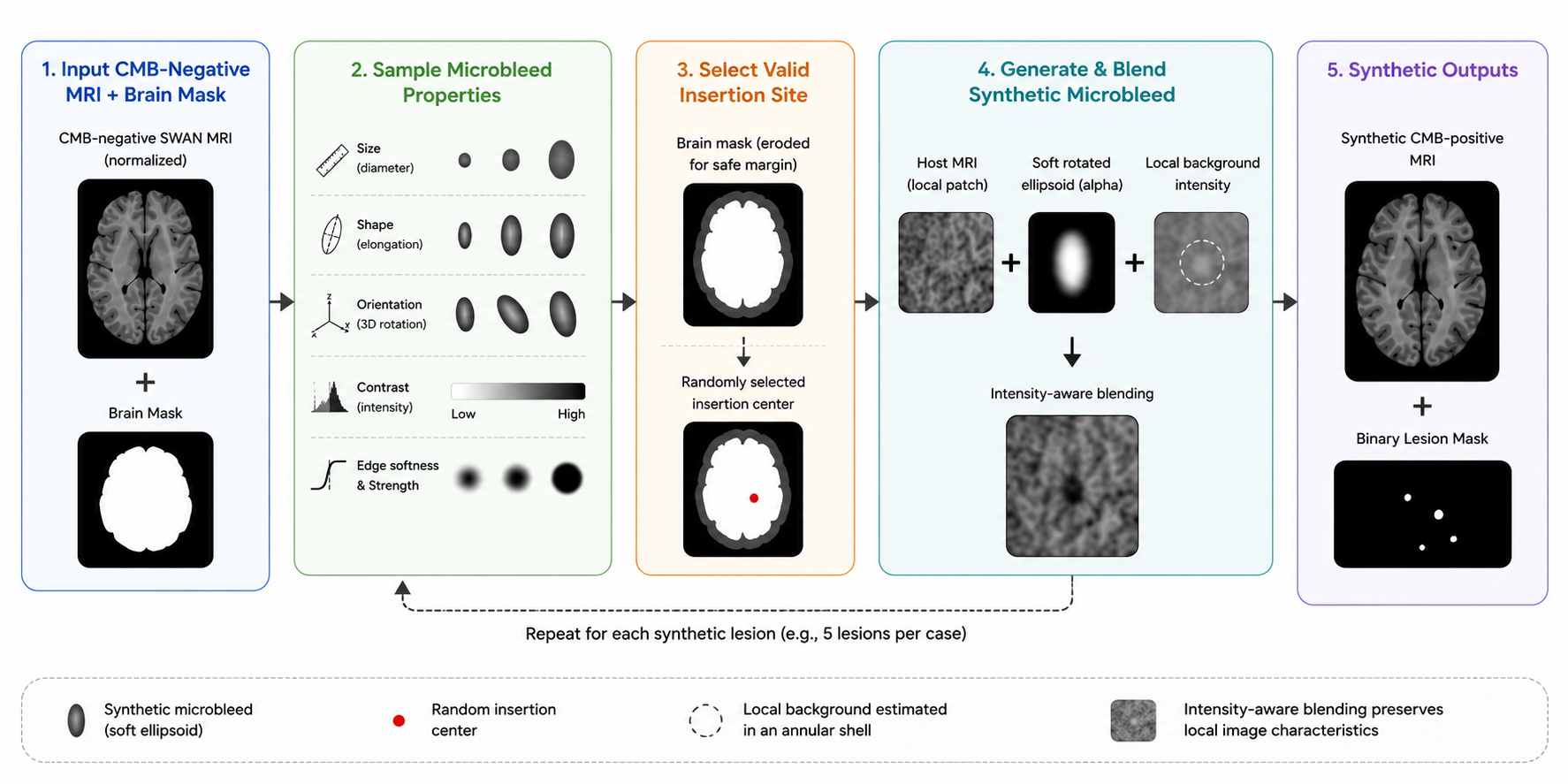} 
\caption{Synthetic CMB Generation Pipeline}
\label{fig:synthetic_CMB_generation_flow}
\end{figure}

The synthetic lesion generation pipeline is illustrated in Fig.~\ref{fig:synthetic_CMB_generation_flow}. Starting from a host CMB-negative SWAN MRI and its corresponding brain mask, synthetic CMBs were sequentially inserted using a parameterized lesion model informed by measurements from the real-positive training cohort. The procedure was repeated independently until the desired lesion burden was reached.

\paragraph{Lesion parameter sampling.}
For each lesion, the equivalent diameter, elongation ratio, signal contrast, and blending strength were sampled from predefined truncated Gaussian distributions (Table~\ref{tab:synthetic_cmb_parameters}). The distributions for lesion size, elongation, and signal contrast were selected with reference to measurements from real CMBs in the training cohort, while the remaining parameters were predefined as part of the synthesis procedure.

As a simplifying geometric assumption, each synthetic CMB was modeled as an ellipsoid. Lesion size and shape were controlled independently by the sampled equivalent diameter \(d\) and elongation ratio \(q\), respectively. The equivalent diameter was used to define the target lesion volume as the volume of a sphere with diameter \(d\),

\begin{equation}
V = \frac{4}{3}\pi\left(\frac{d}{2}\right)^3.
\end{equation}

The ellipsoid was then defined by radii \(r_x=r_y=r\) and \(r_z=qr\). To introduce the sampled elongation while preserving the target lesion volume, the minor-axis radius \(r\) was calculated as

\begin{equation}
r=\left(\frac{3V}{4\pi q}\right)^{1/3}.
\end{equation}

Thus, \(d\) determined the overall lesion size, whereas \(q\) controlled its elongation without altering the target volume. The resulting ellipsoid was randomly rotated in three-dimensional space to generate variable orientations.

\begin{table}[htbp]
\centering
\caption{Parameters used for synthetic cerebral microbleed generation. Selected parameters were informed by measurements from real CMBs in the training cohort, while others were predefined as part of the synthesis procedure.}
\label{tab:synthetic_cmb_parameters}
\small
\setlength{\tabcolsep}{12pt}
\begin{tabular}{lll}
\toprule
\textbf{Parameter} &
\textbf{Sampling / Value} &
\textbf{Range} \\
\midrule

Lesions per scan
& Fixed
& 5 \\

Equivalent diameter $d$
& $\mathcal{N}(3.6,\,0.8^2)$ mm
& 2.0--6.0 mm \\

Elongation ratio $q$
& $\mathcal{N}(1.6,\,0.35^2)$
& 1.0--2.5 \\

Signal contrast $c$
& $\mathcal{N}(160,\,30^2)$
& 90--230 \\

Blending strength $s$
& $\mathcal{N}(0.95,\,0.05^2)$
& 0.85--1.00 \\

Orientation
& Random 3D rotation
& -- \\

Synthetic lesion intensity $I_{\mathrm{lesion}}$
& $I_{\mathrm{bg}}-\mathrm{c}$
& Clipped to 0--60 \\

Binary mask threshold
& Fixed
& $\alpha > 0.5$ \\

\bottomrule
\end{tabular}
\end{table}

\begin{figure}[htbp]
\centering
\includegraphics[width=\textwidth]{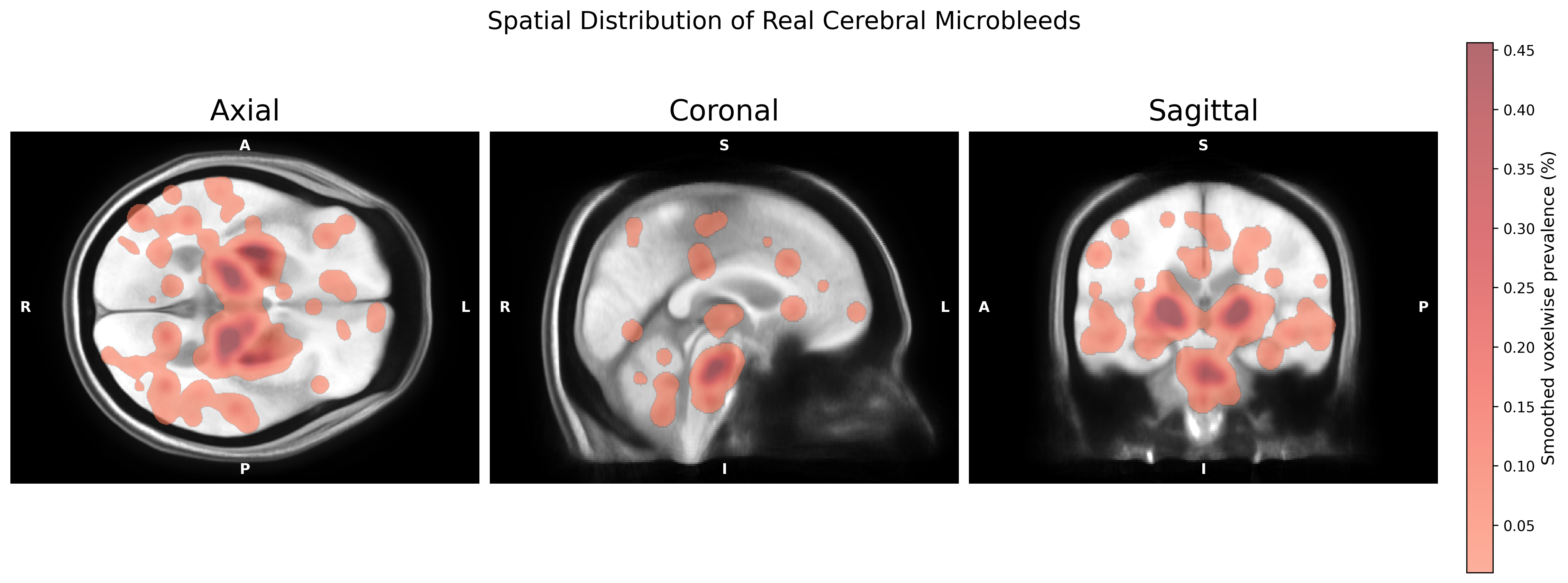} 
\caption{\textbf{Spatial distribution of real CMBs in the training cohort.} Registered lesion masks from CMB-positive training subjects were aggregated into a voxelwise prevalence map for visualization. Although regional differences in lesion prevalence were observed, CMBs occurred throughout much of the brain, supporting the use of uniform lesion insertion within the brain mask during synthetic data generation.}
\label{fig:cmb_heatmap}
\end{figure}

\paragraph{Insertion site selection.}
Lesion centers were sampled uniformly from an eroded brain mask to ensure complete containment within the brain parenchyma while avoiding boundary artifacts. Because cerebral microbleeds were observed across much of the brain rather than being confined to a few specific anatomical regions (Fig.~\ref{fig:cmb_heatmap}), we did not impose an empirical location prior during synthesis. This design was intended to reduce reliance on fixed spatial priors during training. Five lesions were inserted into each healthy MRI, approximately matching the median lesion burden of the training cohort (Fig.~\ref{fig:real_vs_synthetic_characteristics_composite}(a)).

\paragraph{Lesion synthesis and blending.}
After generating the ellipsoidal lesion geometry, a soft alpha mask \(\alpha \in [0,1]\) with sigmoid boundaries was created to provide a smooth transition between the lesion and surrounding tissue. The local background intensity, \(I_{\mathrm{bg}}\), was estimated as the median intensity within a 3--6~mm spherical shell surrounding the insertion site. The target lesion intensity, \(I_{\mathrm{lesion}}\), was then obtained by subtracting the sampled signal contrast from \(I_{\mathrm{bg}}\).

The synthetic lesion was blended into the host image using the soft alpha mask, scaled by the sampled blending strength \(s\), such that \(a=s\alpha\):
\begin{equation}
I_{\mathrm{syn}} = (1-a)I_{\mathrm{host}}
+
aI_{\mathrm{lesion}}.
\end{equation}
This produces a hypointense lesion with a gradual transition to the surrounding tissue. The corresponding binary lesion mask used for training was obtained by thresholding the unscaled alpha mask at \(\alpha > 0.5\). Additional implementation parameters are summarized in Table~\ref{tab:synthetic_cmb_parameters}.

\subsubsection{Comparison of Real and Synthetic Lesion Distribution}

\begin{figure}[htbp]
    \centering
    \includegraphics[width=\textwidth]{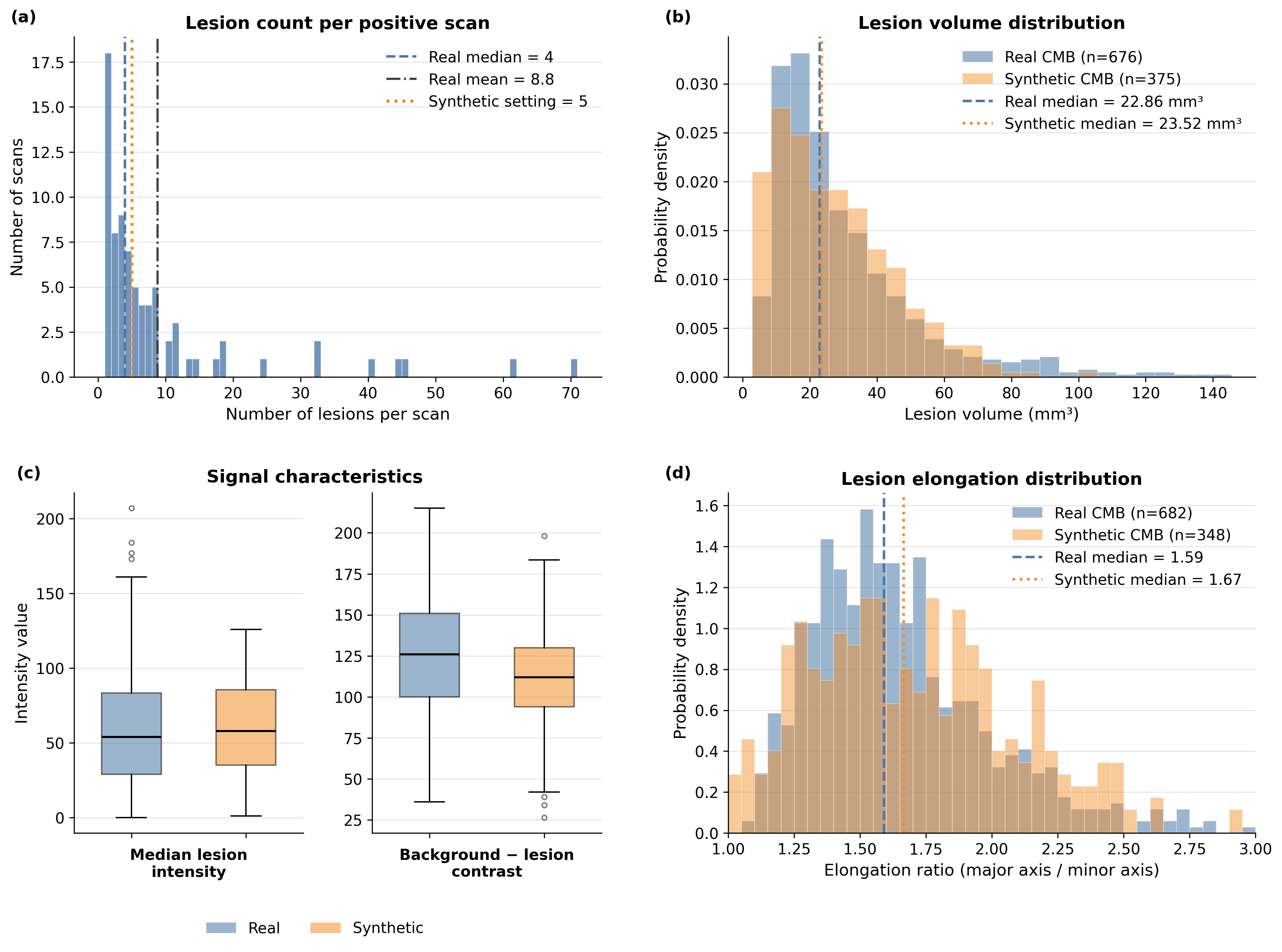}
    \caption{
        Comparison of real and synthetic cerebral microbleed characteristics. Sample sizes may differ across panels because extreme values outside the visualization ranges were excluded from plotting; summary statistics were calculated from the full valid distributions unless otherwise specified.
        (a) Lesion count per positive MRI examination, with five lesions inserted into each synthetic volume.
        (b) Lesion volume distributions.
        (c) Median lesion intensity and background-to-lesion contrast.
        (d) Lesion elongation ratio distributions, defined as the ratio of the major to minor axis length.
    }
    \label{fig:real_vs_synthetic_characteristics_composite}
\end{figure}

\begin{figure}[htbp]
\centering
\includegraphics[width=\textwidth]{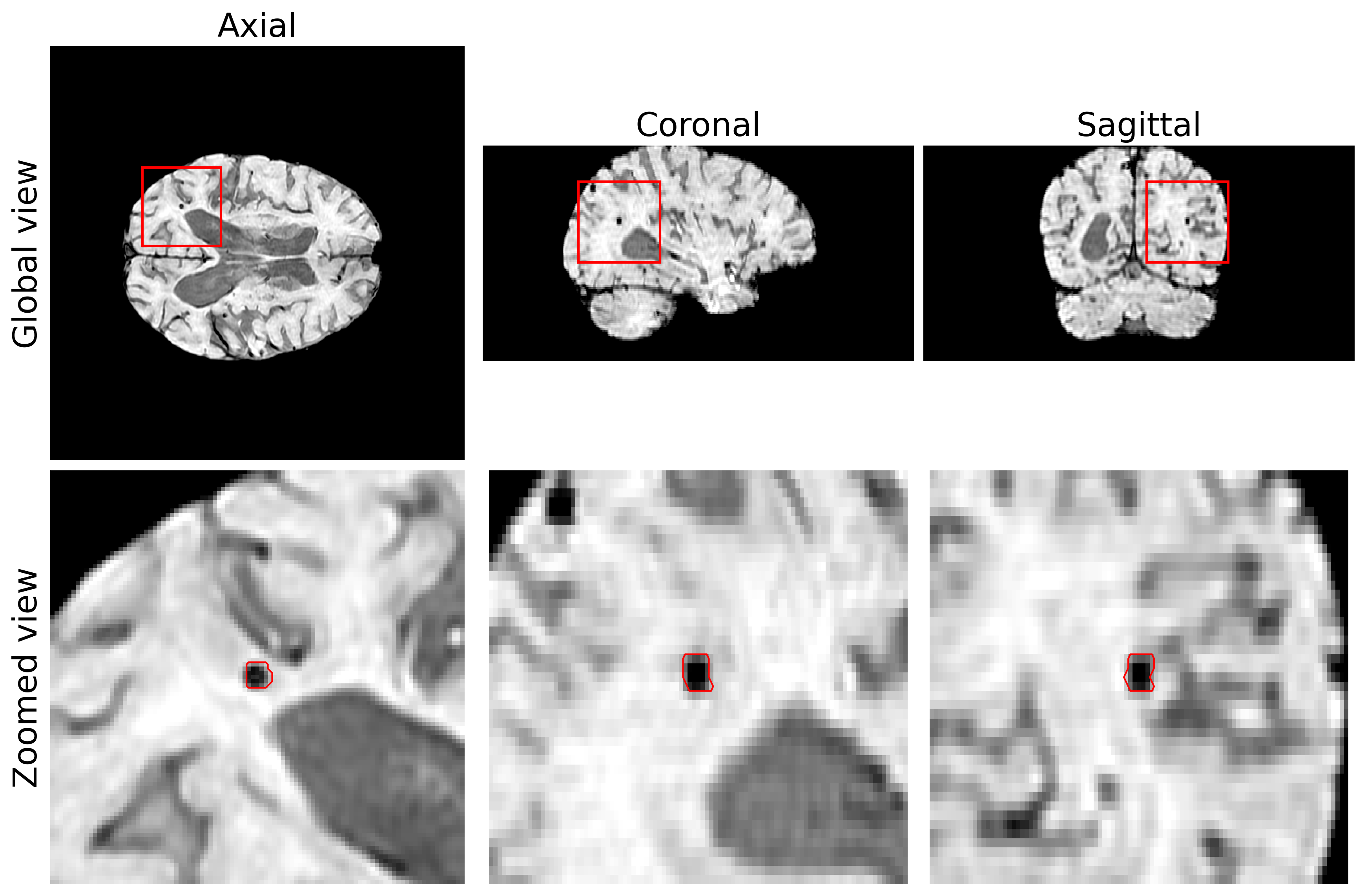} 
\caption{Real CMB Example}
\label{fig:realCMB_sample}
\end{figure}

\begin{figure}[htbp]
\centering
\includegraphics[width=\textwidth]{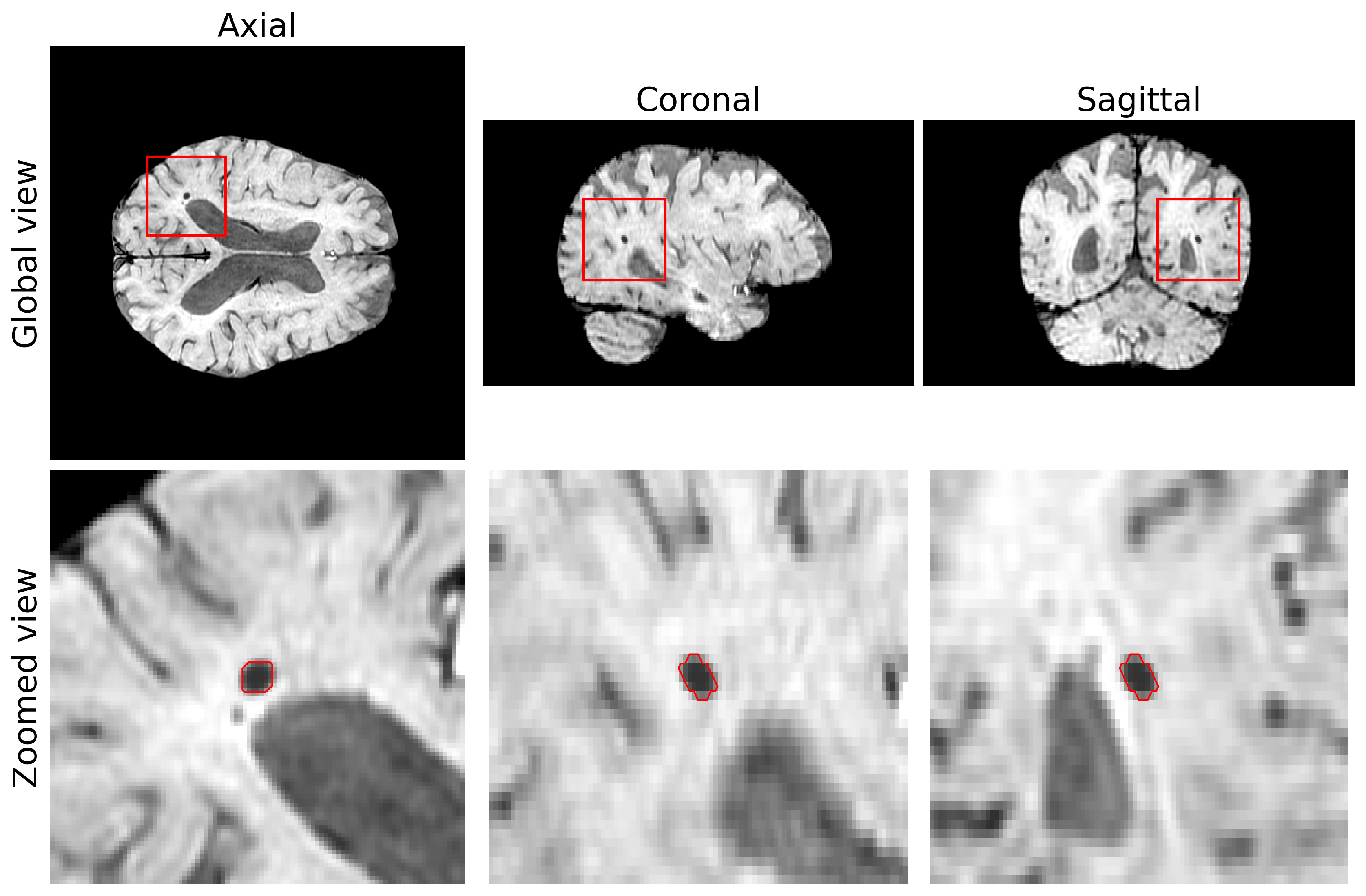} 
\caption{Synthetic CMB Example}
\label{fig:synthCMB_sample}
\end{figure}

Figure~\ref{fig:real_vs_synthetic_characteristics_composite} compares lesion burden, volume, signal characteristics, and elongation between real and synthetic CMBs. As the synthetic generator was parameterized with reference to selected characteristics of real CMBs in the training cohort, the resulting synthetic lesions were designed to approximate, rather than directly reproduce, the corresponding real-lesion distributions. Visualization of real and synthetic CMBs can be found in Fig.~\ref{fig:realCMB_sample} and Fig.~\ref{fig:synthCMB_sample}.

Table~\ref{tab:real_vs_synthetic_characteristics} summarizes these comparisons using the median and interquartile range (IQR), together with the two-sample Kolmogorov-Smirnov statistic (D) to quantify the maximum difference between the empirical cumulative distributions. Overall, the synthetic lesions reproduced the central tendency and distributional characteristics of lesion volume and median intensity relatively closely, while moderate differences remained in elongation and background-to-lesion contrast. These comparisons characterize the degree of similarity achieved by the generator rather than implying exact distributional equivalence between real and synthetic lesions.

\begin{figure}[htbp]
    \centering
    \includegraphics[width=\textwidth]{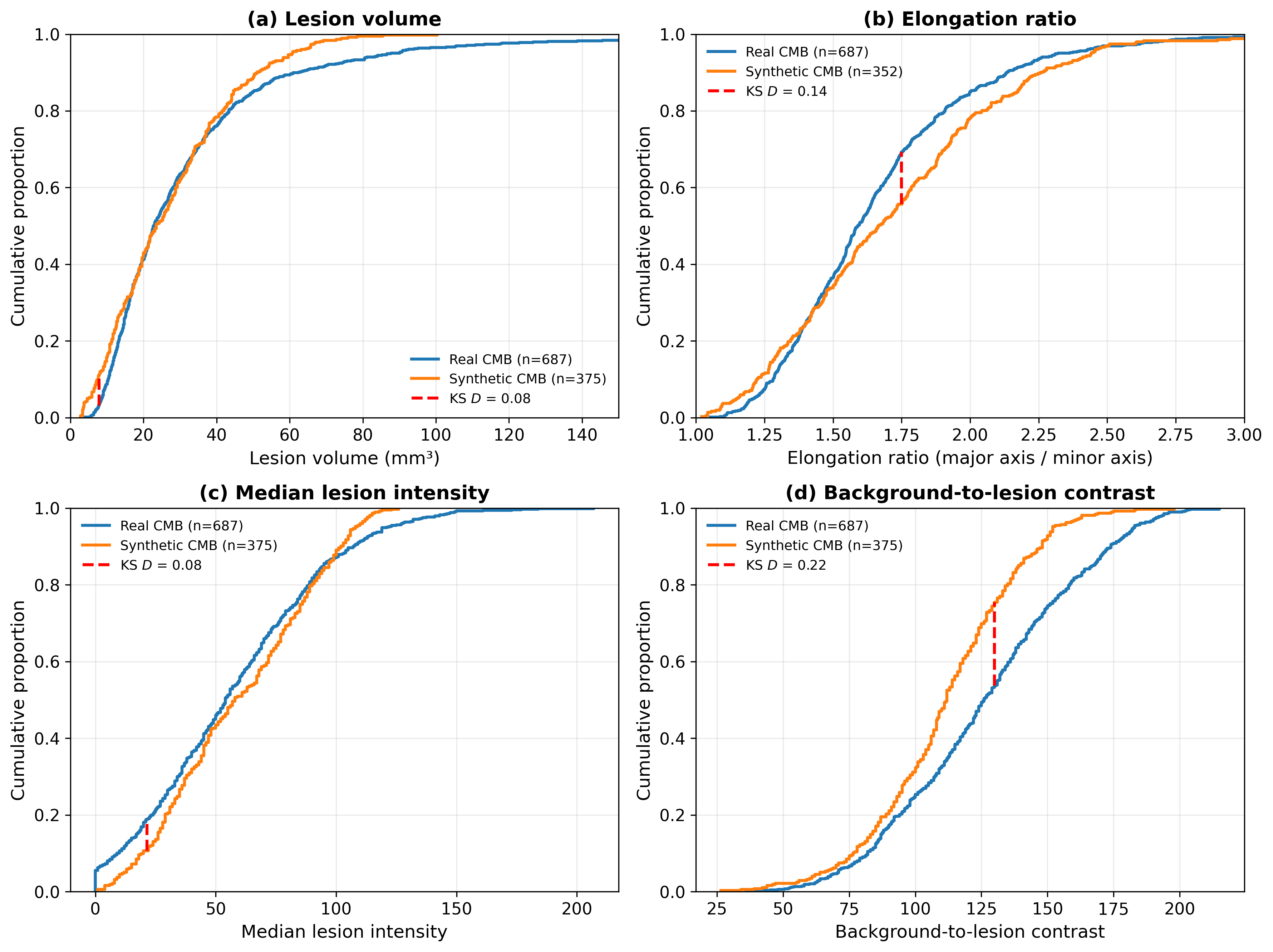}
    \caption{
        Empirical cumulative distribution function (ECDF) comparisons of real and synthetic cerebral microbleed characteristics. (a) Lesion volume. (b) Elongation ratio, defined as the ratio of the major to minor axis length. (c) Median lesion intensity. (d) Background-to-lesion contrast. The red dashed segment in each panel indicates the maximum difference between the real and synthetic ECDFs, corresponding to the two-sample Kolmogorov--Smirnov statistic ($D$). KS statistics were calculated using the full valid distributions; axis limits are used only for visualization where applicable.
    }
    \label{fig:real_vs_synthetic_ecdf_composite}
\end{figure}

\begin{table}[htbp]
\centering
\caption{
    Comparison of real and synthetic cerebral microbleed characteristics.
    Values are reported as median (IQR). The Kolmogorov--Smirnov statistic
    ($D$) quantifies the maximum difference between the empirical cumulative
    distributions of real and synthetic lesions, with corresponding
    two-sided $p$-values.
}
\label{tab:real_vs_synthetic_characteristics}
\small
\begin{tabular}{lcccc}
\toprule
\textbf{Characteristic} &
\textbf{Real CMB} &
\textbf{Synthetic CMB} &
\textbf{KS $D$} &
\textbf{$p$-value} \\
\midrule

Lesion volume (mm$^3$)
& 22.86 (14.73--38.46)
& 23.52 (12.79--37.58)
& 0.08
& 0.084 \\

Elongation ratio
& 1.59 (1.40--1.83)
& 1.67 (1.41--1.95)
& 0.14
& $3.16 \times 10^{-4}$ \\

Median lesion intensity
& 54.00 (29.00--83.25)
& 58.00 (35.00--85.50)
& 0.08
& 0.068 \\

Background-to-lesion contrast
& 126.00 (100.00--151.00)
& 112.00 (94.00--130.00)
& 0.22
& $1.50 \times 10^{-10}$ \\

\bottomrule
\end{tabular}
\end{table}

\subsection{Detection Network}
\label{subsec:detection_network}
CMB detection was formulated as a three-dimensional voxel-wise segmentation task. We evaluated two segmentation frameworks: a configurable 3D U-Net implemented using MONAI~\cite{cardoso2022monai} and the self-configuring nnU-Net framework~\cite{isensee2021nnunet}. The MONAI 3D U-Net was adopted as the primary detection network throughout all experiments, providing a fixed architecture for evaluating different training datasets under a consistent experimental setting. The nnU-Net was included as a complementary baseline to investigate whether the effectiveness of synthetic-to-real transfer depends on the segmentation framework and training pipeline.
\subsubsection{MONAI 3D U-Net}
\label{subsec:monai_3d_unet}
The MONAI detection network consisted of a lightweight three-dimensional U-Net with three resolution levels. The encoder employed channel widths of 16, 32, and 64 feature maps with two downsampling stages, while the decoder followed the corresponding symmetric architecture with residual units at each level. A single-channel SWAN image was used as input, and the network produced a voxel-wise probability map indicating the likelihood of cerebral microbleed presence.

During training, image intensities were linearly normalized to the range [0,1], followed by foreground cropping to remove background voxels. To address the extreme class imbalance between lesion and non-lesion voxels, training patches of size $(96 \times 96 \times 48)$ were sampled using MONAI's positive/negative patch sampler with a positive-to-negative ratio of 2:1. Random flipping along all three spatial axes and random $90^\circ $rotations were applied as data augmentation. The network was optimized using the Dice loss with the Adam optimizer and a cosine annealing learning-rate schedule. Unless otherwise specified, models were trained for 150 epochs.

During inference, full-volume prediction was performed using Gaussian-weighted sliding-window inference with the same patch size as training. Mirror test-time augmentation was applied by averaging predictions over flipped image orientations. The resulting probability map was thresholded at 0.5 to obtain a binary segmentation mask, followed by connected-component analysis to remove predicted components smaller than 15 voxels. Binary segmentations from the five cross-validation models were then combined using voxel-wise majority voting to produce the final prediction for each test case, and all reported performance metrics were computed from these ensembled predictions.

\subsubsection{nnU-Net}
nnU-Net v2 was included as a complementary segmentation framework. nnU-Net is a widely adopted self-configuring framework for biomedical image segmentation that automatically adapts its preprocessing, network architecture, training configuration, and inference strategy to the characteristics of a given dataset. To preserve the self-configuring design philosophy of nnU-Net, no modifications were made to its default settings. We used the three-dimensional full-resolution configuration and retained the automatically determined network and training settings without manual architectural tuning. For consistency with the MONAI experiments, binary segmentations from the five cross-validation models were combined using voxel-wise majority voting to produce the final prediction for each test case, and all reported performance metrics were computed from these final predictions. The same nnU-Net v2 pipeline was applied to models trained on real and synthetic datasets, allowing us to examine whether synthetic-to-real transfer was consistent across segmentation frameworks. 

\subsection{Training Paradigms}
\label{subsec:training_paradigms}

Models were evaluated under real, synthetic, and hybrid training settings. For the data-efficiency experiments, nested real-data subsets of 20, 40, and 78 examinations (R20, R40, and R78) and synthetic-data subsets of 20, 40, and 75 examinations (S20, S40, and S75) were used. Two additional hybrid settings were evaluated: (1) Synth-Train Real-Val, in which models were trained on S75 with real examinations used for validation and checkpoint selection, and (2) mixed training, in which real and synthetic examinations were combined at approximately a 1:1 ratio (R40+S40 and R78+S75).

\subsection{Evaluation Metrics}
Model performance was evaluated from voxel-, lesion-, and case-level perspectives on the independent hold-out test set. Unless otherwise specified, connected-component analysis was performed using 26-connectivity, and components smaller than 15 voxels were removed prior to lesion- and case-level evaluation.
\paragraph{Voxel-wise segmentation performance}
Voxel-wise segmentation accuracy was evaluated using the Dice similarity coefficient (Dice). Because Dice is not informative for CMB-negative test subjects, the mean Dice score was computed over the 26 CMB-positive test subjects.
\paragraph{Lesion-wise detection performance}
Lesion-level performance was evaluated using lesion sensitivity and the number of false-positive lesions per case (FP/case). A ground-truth lesion was considered successfully detected if it exhibited any voxel overlap with a predicted lesion. Lesion sensitivity was defined as the proportion of all ground-truth lesions detected across the hold-out test set, while FP/case was computed as the average number of predicted lesions without overlap with any ground-truth lesion.
\paragraph{Case-level classification performance}
To assess the clinical utility of the detector as a screening tool, voxel-wise segmentation results were further converted into binary case-level predictions. A subject was classified as CMB-positive if at least one predicted lesion remained after connected-component filtering; otherwise, the subject was classified as CMB-negative. Case-level performance was summarized using the F1 score, which balances sensitivity and precision for binary classification.

\section{Results}
The following experiments were designed to investigate the practical utility of synthetic CMBs for training detection models. All models were evaluated on the same independent hold-out test set. We examined four questions: (1) whether models trained exclusively on synthetic CMBs can generalize to real CMB detection; (2) how synthetic and real training data compare across different training set sizes; (3) whether incorporating limited real data through validation or mixed training further improves synthetic-to-real performance; and (4) whether synthetic-to-real transfer is consistent across different detection architectures. 

Voxel- and lesion-level performance across all experiments is summarized in Table~\ref{tab:all_lesion_metrics}, while case-level classification performance, obtained by considering any predicted CMB as a positive prediction, is reported in Table~\ref{tab:all_case_metrics}. Stratified lesion sensitivity based on different CMB volume sizes is reported in Table~\ref{tab:stratified_lesion_sensitivity}.

For consistency, experiments are named according to the detection framework and training data composition. $R$ and $S$ indicate real and synthetic training examinations, respectively, followed by the number of examinations used (e.g., MONAI R78 and nnUNet S75). Combined datasets are denoted using $+$ (e.g., MONAI R40+S40), while S75+ReVal denotes synthetic-only training with real CMB examinations used exclusively for validation and checkpoint selection. MONAI experiments use the 3D U-Net framework described in Section~\ref{subsec:monai_3d_unet}.

\subsection{Pure Synthetic-To-Real Transfer}
We first investigated whether a model trained exclusively on synthetic CMBs could generalize to real CMB detection. MONAI S75 was trained on 75 CMB-negative examinations with synthetic lesions, without exposure to any real CMB annotations during either training or validation. MONAI R78, trained on 78 examinations containing real annotated CMBs under the same training framework, served as the real-data reference.

As shown in Table~\ref{tab:real_vs_synth}, MONAI S75 retained 83.7\% of the mean Dice and 87.7\% of the lesion sensitivity achieved by MONAI R78, despite being trained without real CMB annotations. This transfer was accompanied by a higher false-positive rate (1.423 vs. 1.134 FP/case).

\begin{figure}[htbp]
\centering
\includegraphics[width=\textwidth]{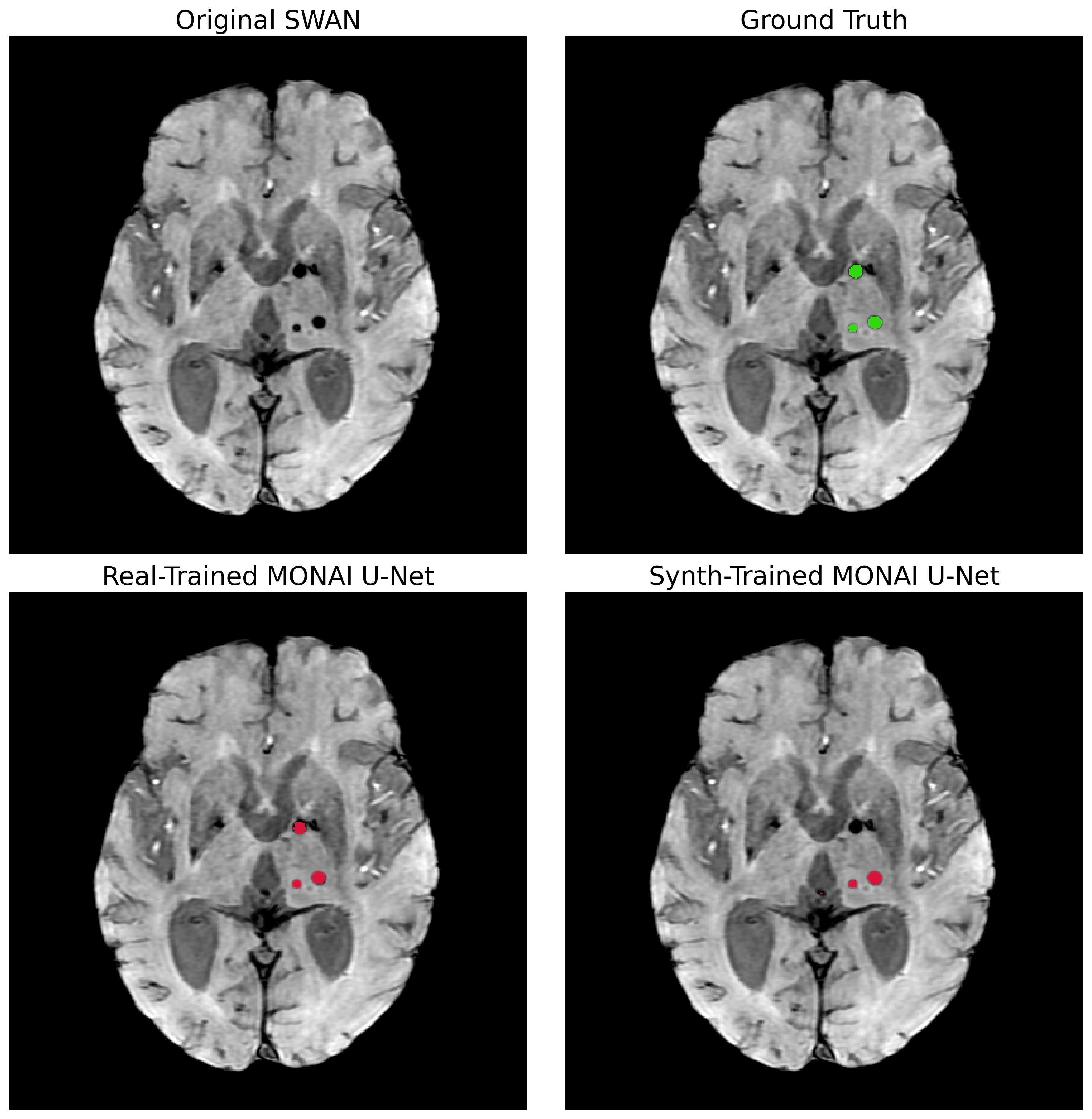} 
\caption{Prediction example of real-trained model and synth-trained model}
\label{fig:synth_vs_real_prediction}
\end{figure}

Nevertheless, MONAI S75 retained substantial detection and segmentation performance when evaluated on real CMBs. Representative predictions from MONAI R78 and MONAI S75 are shown in Fig.~\ref{fig:synth_vs_real_prediction}.

\begin{table}[htbp]
\centering
\caption{Performance of MONAI 3D U-Net models trained on real (R78) and synthetic (S75) data on the independent hold-out test set. Values are reported as point estimates with 95\% bootstrap confidence intervals. Higher values indicate better performance except for FP/Case. The final column reports the performance of the synthetic model relative to the real-data model.}
\label{tab:real_vs_synth}
\small
\begin{tabular}{lccc}
\toprule
\textbf{Metric} & \textbf{MONAI R78} & \textbf{MONAI S75} & \textbf{Relative (\%)} \\
\midrule
Mean Dice &
0.688 (0.621--0.749) &
0.576 (0.508--0.641) &
83.7 \\

Lesion Sensitivity &
0.854 (0.797--0.910) &
0.749 (0.656--0.830) &
87.7 \\

FP/Case $\downarrow$ &
1.135 (0.769--1.538) &
1.423 (1.077--1.789) &
125.4 \\

Lesion Precision &
0.716 (0.601--0.798) &
0.636 (0.526--0.711) &
88.7 \\

Lesion F1 &
0.779 (0.699--0.835) &
0.687 (0.596--0.752) &
88.2 \\

\midrule

Case Sensitivity &
1.000 (1.000--1.000) &
1.000 (1.000--1.000) &
100.0 \\

Case Specificity &
0.577 (0.385--0.769) &
0.308 (0.154--0.500) &
53.3 \\

Case Precision &
0.703 (0.619--0.812) &
0.591 (0.542--0.667) &
84.1 \\

Case F1 &
0.825 (0.765--0.897) &
0.743 (0.703--0.800) &
90.0 \\
\bottomrule
\end{tabular}
\end{table}

\subsection{Effect of Training Set Size}

We next examined the effect of training set size for real and synthetic CMB training. MONAI models were trained using increasing numbers of real (R20, R40, and R78) or synthetic (S20, S40, and S75) examinations, as described in Section~\ref{subsec:training_paradigms}. The training protocol was otherwise identical, except that R20 and S20 were trained for 300 epochs instead of 150 to allow sufficient convergence with the smaller training sets.

As shown in the ``Effect of Training Set Size'' group in Table~\ref{tab:all_lesion_metrics}, real-data training produced higher Dice and lesion sensitivity than synthetic-only training across the evaluated training set sizes. Notably, MONAI R20, trained using only 20 real CMB-positive examinations, achieved a mean Dice of 0.660 and lesion sensitivity of 0.807, exceeding all three synthetic-only configurations, including MONAI S75 (Dice: 0.576; lesion sensitivity: 0.749). 

These results show that the performance gap between real- and synthetic-data training persisted across the evaluated training set sizes.

\subsection{Incorporating Real and Synthetic Data}
We further evaluated whether combining data from real and synthetic CMB data could improve performance over single-source training. Two strategies were considered: using real CMB examinations exclusively for validation and checkpoint selection during synthetic training (S75+ReVal), and directly combining real and synthetic examinations during training (R40+S40 and R78+S75).

Using real data solely for validation resulted in only modest changes relative to pure synthetic training. Compared with MONAI S75, S75+ReVal increased mean Dice from 0.576 to 0.585 and lesion sensitivity from 0.749 to 0.760, while FP/case increased slightly from 1.423 to 1.462. The confidence intervals substantially overlapped across all three metrics, indicating no clear overall performance improvement from incorporating real data solely for model selection.

When synthetic examinations were added directly to real-data training, both R40+S40 and R78+S75 showed higher lesion sensitivity than their real-only counterparts, accompanied by lower Dice and higher FP/case (Table~\ref{tab:all_lesion_metrics}). The observed trade-off was more pronounced for R78+S75, where sensitivity increased from 0.854 to 0.877 while FP/case increased from 1.135 to 1.788. Overall, synthetic augmentation did not produce a clear improvement across metrics.

\subsection{Synthetic-to-Real Transfer Across Detection Frameworks}
\begin{figure}[htbp]
\centering
\includegraphics[width=\textwidth]{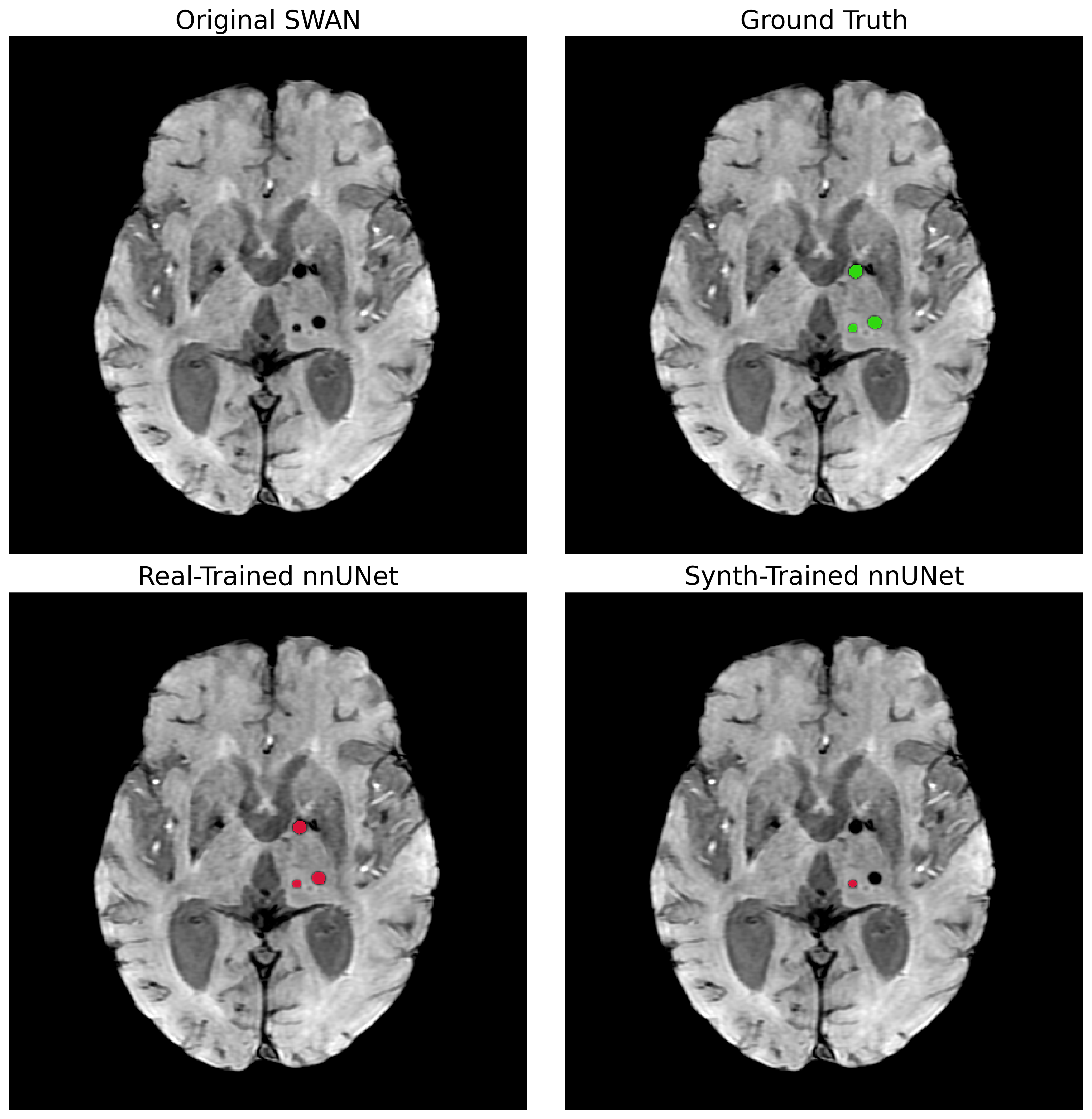} 
\caption{The synth-trained nnU-Net produced substantially fewer predicted lesion voxels, consistent with its lower lesion sensitivity.}
\label{fig:nnUNet_prediction}
\end{figure}

Finally, we investigated whether synthetic-to-real transfer was consistent across detection frameworks by comparing real-only and synthetic-only training using both MONAI 3D U-Net and nnUNet. Results are summarized in the ''Model Dependency'' group of Table~\ref{tab:all_lesion_metrics}, with representative predictions shown in Fig.~\ref{fig:synth_vs_real_prediction} and Fig.~\ref{fig:nnUNet_prediction}.

The two frameworks showed different synthetic-to-real transfer. As reported above, MONAI S75 retained substantial performance relative to MONAI R78, with mean Dice decreasing from 0.688 to 0.576 and lesion sensitivity from 0.854 to 0.749. In contrast, nnUNet achieved the strongest real-data performance among the evaluated models (Dice: 0.777; lesion sensitivity: 0.860), but performance decreased substantially under synthetic-only training, with nnUNet S75 achieving a Dice of 0.240 and lesion sensitivity of 0.269. FP/case simultaneously decreased from 0.500 to 0.192.

These results demonstrated a notable difference in synthetic-to-real transfer between the two evaluated detection frameworks.

\begin{table}[htbp]
\centering
\caption{Performance of all experiments on the independent hold-out test set. Values are reported as point estimates with 95\% bootstrap confidence intervals. Higher values indicate better performance except for FP/Case. Rows shaded in gray indicate the reference experiments (MONAI R78 and MONAI S75), which are intentionally repeated across experiment groups to facilitate direct comparison.}
\label{tab:all_lesion_metrics}
\small
\setlength{\tabcolsep}{3.5pt}

\begin{tabular}{lccc}
\toprule
\textbf{Experiment} &
\textbf{Mean Dice} &
\textbf{Lesion Sensitivity} &
\textbf{FP/Case $\downarrow$} \\
\midrule

\multicolumn{4}{l}{\textbf{Pure Synthetic-To-Real Transfer}}\\
\addlinespace[2pt]

\rowcolor{gray!10}
MONAI R78
& \underline{0.688 (0.621--0.749)}
& \underline{0.854 (0.797--0.910)}
& \underline{1.135 (0.769--1.538)} \\

\rowcolor{gray!10}
MONAI S75
& 0.576 (0.508--0.641)
& 0.749 (0.656--0.830)
& 1.423 (1.077--1.788) \\

\midrule

\multicolumn{4}{l}{\textbf{Effect of Training Set Size}}\\
\addlinespace[2pt]

MONAI R20
& 0.660 (0.592--0.724)
& 0.807 (0.733--0.881)
& 1.385 (0.923--1.923) \\

MONAI R40
& 0.655 (0.592--0.714)
& 0.819 (0.747--0.889)
& 1.269 (0.846--1.750) \\

\rowcolor{gray!10}
MONAI R78
& \underline{0.688 (0.621--0.749)}
& \underline{0.854 (0.797--0.910)}
& \underline{1.135 (0.769--1.538)} \\

MONAI S20
& 0.587 (0.516--0.656)
& 0.737 (0.632--0.833)
& \underline{1.173 (0.827--1.538)} \\

MONAI S40
& \underline{0.599 (0.540--0.658)}
& \underline{0.754 (0.657--0.851)}
& 1.596 (1.231--2.000) \\

\rowcolor{gray!10}
MONAI S75
& 0.576 (0.508--0.641)
& 0.749 (0.656--0.830)
& 1.423 (1.077--1.788) \\

\midrule

\multicolumn{4}{l}{\textbf{Mixed Training}}\\
\addlinespace[2pt]

\rowcolor{gray!10}
MONAI S75
& 0.576 (0.508--0.641)
& 0.749 (0.656--0.830)
& 1.423 (1.077--1.788) \\

MONAI S75+ReVal
& 0.585 (0.522--0.647)
& 0.760 (0.674--0.842)
& 1.462 (1.096--1.846) \\

MONAI R40+S40
& 0.649 (0.576--0.719)
& 0.830 (0.761--0.896)
& 1.596 (1.135--2.096) \\

MONAI R78+S75
& 0.656 (0.584--0.723)
& \underline{0.877 (0.817--0.924)}
& 1.788 (1.269--2.346) \\

\rowcolor{gray!10}
MONAI R78
& \underline{0.688 (0.621--0.749)}
& 0.854 (0.797--0.910)
& \underline{1.135 (0.769--1.538)} \\

\midrule

\multicolumn{4}{l}{\textbf{Model Dependency}}\\
\addlinespace[2pt]

\rowcolor{gray!10}
MONAI R78
& 0.688 (0.621--0.749)
& 0.854 (0.797--0.910)
& 1.135 (0.769--1.538) \\

\rowcolor{gray!10}
MONAI S75
& 0.576 (0.508--0.641)
& 0.749 (0.656--0.830)
& 1.423 (1.077--1.788) \\

nnUNet R78
& \underline{0.777 (0.719--0.828)}
& \underline{0.860 (0.801--0.926)}
& 0.500 (0.269--0.769) \\

nnUNet S75
& 0.240 (0.152--0.341)
& 0.269 (0.195--0.349)
& \underline{0.192 (0.096--0.308)} \\

\bottomrule
\end{tabular}
\end{table}

\begin{table}[htbp]
\centering
\caption{Case-level performance of all experiments on the independent hold-out test set. Values are reported as point estimates with 95\% bootstrap confidence intervals. Higher values indicate better performance. Rows shaded in gray indicate the reference experiments (MONAI Real78 and MONAI Synth75), which are intentionally repeated across experiment groups to facilitate direct comparison.}
\label{tab:all_case_metrics}
\small
\setlength{\tabcolsep}{3.5pt}

\begin{tabular}{lccc}
\toprule
\textbf{Experiment} &
\textbf{Case Sensitivity} &
\textbf{Case Specificity} &
\textbf{Case F1} \\
\midrule

\multicolumn{4}{l}{\textbf{Pure Synthetic vs. Real}}\\
\addlinespace[2pt]

\rowcolor{gray!10}
MONAI R78
& \underline{1.000 (1.000--1.000)}
& \underline{0.577 (0.385--0.769)}
& \underline{0.825 (0.765--0.897)} \\

\rowcolor{gray!10}
MONAI S75
& 1.000 (1.000--1.000)
& 0.308 (0.154--0.500)
& 0.743 (0.703--0.800) \\

\midrule

\multicolumn{4}{l}{\textbf{Effect of Training Set Size}}\\
\addlinespace[2pt]

MONAI R20
& 1.000 (1.000--1.000)
& 0.462 (0.269--0.654)
& 0.788 (0.732--0.852) \\

MONAI R40
& 1.000 (1.000--1.000)
& 0.500 (0.308--0.692)
& 0.800 (0.743--0.867) \\

\rowcolor{gray!10}
MONAI R78
& \underline{1.000 (1.000--1.000)}
& \underline{0.577 (0.385--0.769)}
& \underline{0.825 (0.765--0.897)} \\

MONAI S20
& 1.000 (1.000--1.000)
& 0.462 (0.269--0.654)
& \underline{0.788 (0.732--0.852)} \\

MONAI S40
& 1.000 (1.000--1.000)
& 0.192 (0.077--0.346)
& 0.712 (0.684--0.754) \\

\rowcolor{gray!10}
MONAI S75
& 1.000 (1.000--1.000)
& 0.308 (0.154--0.500)
& 0.743 (0.703--0.800) \\

\midrule

\multicolumn{4}{l}{\textbf{Mixed Training}}\\
\addlinespace[2pt]

\rowcolor{gray!10}
MONAI S75
& 1.000 (1.000--1.000)
& 0.308 (0.154--0.500)
& 0.743 (0.703--0.800) \\

MONAI S75+ReVal
& 1.000 (1.000--1.000)
& 0.385 (0.192--0.577)
& 0.765 (0.712--0.825) \\

MONAI R40+S40
& 1.000 (1.000--1.000)
& 0.385 (0.192--0.577)
& 0.765 (0.712--0.825) \\

MONAI R78+S75
& 1.000 (1.000--1.000)
& 0.385 (0.192--0.577)
& 0.765 (0.712--0.825) \\

\rowcolor{gray!10}
MONAI R78
& \underline{1.000 (1.000--1.000)}
& \underline{0.577 (0.385--0.769)}
& \underline{0.825 (0.765--0.897)} \\

\midrule

\multicolumn{4}{l}{\textbf{Model Dependency}}\\
\addlinespace[2pt]

\rowcolor{gray!10}
MONAI R78
& 1.000 (1.000--1.000)
& 0.577 (0.385--0.769)
& 0.825 (0.765--0.897) \\

\rowcolor{gray!10}
MONAI S75
& \underline{1.000 (1.000--1.000)}
& 0.308 (0.154--0.500)
& 0.743 (0.703--0.800) \\

nnUNet R78
& \underline{1.000 (1.000--1.000)}
& \underline{0.885 (0.731--1.000)}
& \underline{0.945 (0.881--1.000)} \\

nnUNet S75
& 0.846 (0.692--0.962)
& \underline{0.885 (0.731--1.000)}
& \underline{0.863 (0.756--0.943)} \\

\bottomrule
\end{tabular}
\end{table}

\begin{table*}[htbp]
\centering
\caption{
Lesion sensitivity stratified by ground-truth lesion volume on the independent hold-out test set.
Values are reported as point estimates with 95\% bootstrap confidence intervals.
Rows shaded in gray indicate the reference experiments (MONAI R78 and MONAI S75),
which are intentionally repeated across experiment groups to facilitate direct comparison.
}
\label{tab:stratified_lesion_sensitivity}
\small
\setlength{\tabcolsep}{3.5pt}

\begin{tabular}{lcccc}
\toprule
\textbf{Experiment} &
\textbf{$<15$ mm$^3$} &
\textbf{15--25 mm$^3$} &
\textbf{25--40 mm$^3$} &
\textbf{$>40$ mm$^3$} \\
\midrule

\multicolumn{5}{l}{\textbf{Pure Synthetic vs. Real}}\\
\addlinespace[2pt]

\rowcolor{gray!10}
MONAI R78
& 0.680 (0.421--0.909)
& 0.871 (0.739--0.968)
& 0.893 (0.742--1.000)
& 0.951 (0.906--1.000) \\

\rowcolor{gray!10}
MONAI S75
& 0.595 (0.357--0.846)
& 0.783 (0.606--0.917)
& 0.817 (0.639--0.970)
& 0.778 (0.672--0.873) \\

\midrule

\multicolumn{5}{l}{\textbf{Effect of Training Set Size}}\\
\addlinespace[2pt]

MONAI R20
& 0.642 (0.381--0.909)
& 0.852 (0.698--0.968)
& 0.892 (0.744--1.000)
& 0.967 (0.927--1.000) \\

MONAI R40
& 0.556 (0.273--0.840)
& 0.808 (0.645--0.938)
& 0.892 (0.744--1.000)
& 0.951 (0.906--1.000) \\

\rowcolor{gray!10}
MONAI R78
& 0.680 (0.421--0.909)
& 0.871 (0.739--0.968)
& 0.893 (0.742--1.000)
& 0.951 (0.906--1.000) \\

MONAI S20
& 0.556 (0.273--0.840)
& 0.740 (0.565--0.887)
& 0.817 (0.639--0.970)
& 0.812 (0.727--0.890) \\

MONAI S40
& 0.514 (0.250--0.792)
& 0.723 (0.533--0.892)
& 0.815 (0.636--0.963)
& 0.882 (0.812--0.951) \\

\rowcolor{gray!10}
MONAI S75
& 0.595 (0.357--0.846)
& 0.783 (0.606--0.917)
& 0.817 (0.639--0.970)
& 0.778 (0.672--0.873) \\

\midrule

\multicolumn{5}{l}{\textbf{Hybrid Training Strategies}}\\
\addlinespace[2pt]

\rowcolor{gray!10}
MONAI S75
& 0.595 (0.357--0.846)
& 0.783 (0.606--0.917)
& 0.817 (0.639--0.970)
& 0.778 (0.672--0.873) \\

MONAI S75+ReVal
& 0.595 (0.357--0.846)
& 0.765 (0.595--0.912)
& 0.867 (0.707--1.000)
& 0.830 (0.733--0.915) \\

MONAI R40+S40
& 0.593 (0.308--0.850)
& 0.784 (0.613--0.920)
& 0.918 (0.765--1.000)
& 0.967 (0.921--1.000) \\

MONAI R78+S75
& 0.677 (0.455--0.895)
& 0.913 (0.806--0.982)
& 0.892 (0.744--1.000)
& 0.967 (0.921--1.000) \\

\rowcolor{gray!10}
MONAI R78
& 0.680 (0.421--0.909)
& 0.871 (0.739--0.968)
& 0.893 (0.742--1.000)
& 0.951 (0.906--1.000) \\

\midrule

\multicolumn{5}{l}{\textbf{Model Dependency}}\\
\addlinespace[2pt]

\rowcolor{gray!10}
MONAI R78
& 0.680 (0.421--0.909)
& 0.871 (0.739--0.968)
& 0.893 (0.742--1.000)
& 0.951 (0.906--1.000) \\

\rowcolor{gray!10}
MONAI S75
& 0.595 (0.357--0.846)
& 0.783 (0.606--0.917)
& 0.817 (0.639--0.970)
& 0.778 (0.672--0.873) \\

nnUNet R78
& 0.687 (0.429--0.944)
& 0.872 (0.750--0.968)
& 0.868 (0.727--0.972)
& 0.950 (0.900--1.000) \\

nnUNet S75
& 0.318 (0.077--0.600)
& 0.291 (0.167--0.433)
& 0.344 (0.167--0.524)
& 0.210 (0.111--0.343) \\

\bottomrule
\end{tabular}
\end{table*}

\FloatBarrier
\section{Discussion}
Our results show that synthetic CMBs can provide useful supervision for training detectors on real images, even when the synthetic lesions do not perfectly reproduce the available real CMB distribution. By inserting synthetic lesions into real CMB-negative scans, our approach combines controlled lesion cues with authentic anatomical backgrounds and allows negative scans to contribute to detector training. Models trained this way showed substantial transfer to real CMBs, demonstrating that a relatively simple parameterized generator can provide useful training signals. However, real-data training consistently outperformed synthetic-data training across the evaluated data scales, indicating that synthetic data did not fully replace real positive cases in our experiments. This gap should not be considered an inherent limit of synthetic training, as we evaluated only one synthetic generation setting and different parameter ranges or generation strategies may produce different results.

\begin{figure}[htbp]
\centering
\includegraphics[width=\textwidth]{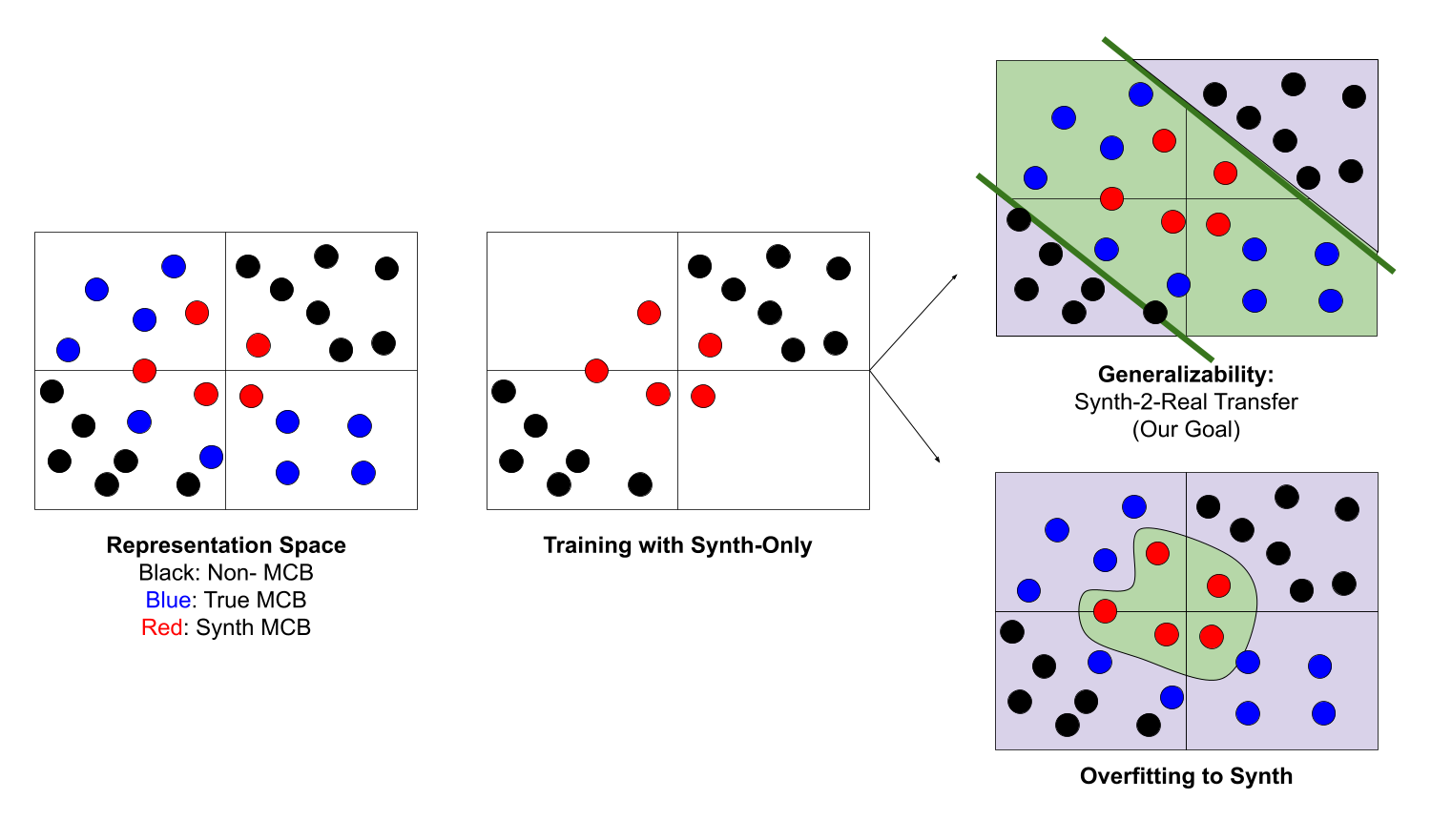} 
\caption{Synthetic to Real Challenge}
\label{fig:synt2real_challenge}
\end{figure}

Interestingly, synthetic-to-real transfer also differed substantially between the two detection frameworks. While the synthetic-trained MONAI 3D U-Net retained substantial performance on real CMBs, the same synthetic training strategy transferred poorly with nnUNet. This suggests that synthetic-to-real transfer depends not only on how closely synthetic lesions resemble real lesions, but also on the downstream learning framework. As conceptually illustrated in Fig.~\ref{fig:synt2real_challenge}, successful transfer may occur when a model learns features that are shared between synthetic and real CMBs, whereas poor transfer may occur when learning becomes more specific to characteristics of the synthetic distribution. However, our experiments do not identify the mechanism behind the difference between MONAI and nnUNet. The two frameworks differ in model capacity, preprocessing, augmentation, and training procedure. Any of these factors may contribute to the observed behavior. Therefore, rather than attributing the difference to a specific component, our results demonstrate that synthetic-to-real transfer is framework-dependent and motivate further investigation into the interaction between synthetic data and downstream learners.

Several limitations should be considered when interpreting these findings. First, CMBs are relatively suitable for parameterized synthesis because they are small, localized lesions with relatively simple visual characteristics and a physics-based explanation for their appearance on susceptibility-sensitive MRI. Unlike larger and more complex lesions such as brain tumors, CMBs also have limited effects on surrounding anatomy, making direct insertion into CMB-negative scans feasible. Therefore, the effectiveness of our approach should not be assumed to generalize to more complex pathologies. Second, our synthetic generator was intentionally simple and modeled only selected CMB characteristics derived from the training cohort. Differences remained between the real and synthetic distributions, particularly in background-to-lesion contrast. Greater variability in lesion shape, boundary appearance, and imaging artifacts could be explored in future generators. Additionally, only one synthetic generation setting was used across the downstream experiments, and different settings could produce different transfer behavior. Third, we evaluated only two downstream frameworks: a lightweight MONAI 3D U-Net and nnUNet. Although their contrasting results were sufficient to demonstrate that synthetic-to-real transfer can vary across detection frameworks, they do not identify which model or training characteristics are responsible for this difference. A more systematic comparison across model capacities and training procedures will be needed to understand the mechanisms underlying this framework dependence. 

Future work should focus not only on generating more realistic synthetic lesions, but also on how synthetic data can improve the practical development of detectors. One important direction is data efficiency: synthetic data may be most useful in extremely low-data settings, where only a small number of real positive cases are available. In our experiments, even models trained with 20 real cases already achieved promising performance, leaving limited room to demonstrate the potential benefit of synthetic data under more severe data scarcity. Synthetic data may also be more useful as a complement to real data rather than as a complete replacement. Its controllability could be used to increase the representation of underrepresented lesion sizes, contrasts, or lesion burdens and to investigate how such changes affect detector behavior. The value of synthetic data should be evaluated by its ability to improve performance on real lesions. Future studies that jointly investigate generator design, real-data availability, and downstream learning frameworks may therefore provide a more practical understanding of when and how synthetic data is most useful for medical image detection.

\bibliographystyle{splncs04}

\end{document}